\documentclass[journal]{IEEEtran} 		

\usepackage{subfig}
\usepackage{graphicx}
\usepackage{amsmath,amssymb,amsfonts}
\usepackage{algorithmic}
\usepackage{graphicx}
\usepackage{multirow}
\usepackage{xcolor}

\newcommand*{\addheight}[2][.5ex]{%
  \raisebox{0pt}[\dimexpr\height+(#1)\relax]{#2}%
}

\title{Video Super-Resolution Using a Grouped Residual in Residual Network}

\author{MohammadHossein Ashoori,~and~Arash Amini,~\IEEEmembership{Senior,~IEEE}\thanks{The authors are with the Department of Electrical Engineering, Sharif University of Technology, Tehran, Iran. (email:mohammadhoseinashoori@gmail.com, aamini@sharif.edu)}}

\begin{document}

\maketitle

\begin{abstract}
Super-resolution (SR) is the technique of increasing the nominal resolution of image / video content accompanied with quality improvement. 
Video super-resolution (VSR) can be considered as the generalization of single image super-resolution (SISR). This generalization should be such that more detail is created in the output using adjacent input frames.
In this paper, we propose a grouped residual in residual network (GRRN) for VSR. By adjusting the hyperparameters of the proposed structure, we train three networks with different numbers of parameters and compare their quantitative and qualitative results with the existing methods.
Although based on some quantitative criteria, GRRN does not provide better results than the existing methods, in terms of the quality of the output image it has acceptable performance.
\end{abstract}

\section{Introduction}

In recent years, the display quality 
of electronic devices such as televisions, cellphones, and laptops has experienced considerable improvements. One of them 
is the capability of displaying high-resolution (HR) content. 
However, much of the available content 
is in the form of low-resolution (LR) digital images and videos. The easiest way to display an LR image/video on an HR display is to apply interpolation methods such as bilinear or bicubic. While these methods increase the number of pixels, they leave the perceived quality of the content almost unchanged.

The purpose of super-resolution (SR) is to generate more pixels with additional details. 
To better clarify the problem of SR, we first consider the opposite task, i.e., converting an HR image/video into the LR format. For this purpose, we need to pass the frames/images 
through a low-pass filter and then, downsample the outcome to the desired size. This implies that high-frequency information are automatically discarded from  the LR images/frames.
 Therefore, a proper SR method essentially requires the recovery of high-frequency information.

In recent years, convolutional neural networks (CNN) have been successful in the single image SR (SISR) task; the SRCNN in \cite{SRCNN} is one of the well-studied examples. 
%
The networks in \cite{Charbonnier, FSRCNN, PixelShuffle} improved the performance of SRCNN \cite{SRCNN} by changing the upsampling module and using slightly deeper structures. 
By using much deeper structures, DBPN \cite{DBPN}, SRResNet \cite{SRGAN}, EDSR \cite{EDSR}, and RCAN \cite{RCAN} marked significantly better final accuracies.
 Similar to ResNet \cite{ResNet}, the last three networks consist of residual blocks. This structure enables us to increase the number of layers without degrading the training procedure. 
SRGAN \cite{SRGAN} and ESRGAN \cite{ESRGAN} are also two other works in the area of SISR that use generative adversarial networks (GAN) to improve the visual quality of the output; note that keeping the training process stable in these types of networks is very difficult.

VSR can be considered as a generalization of SISR. The main difference is that we can use several neighboring input frames to produce a better output. One way to use information from neighboring frames is to use the RNN architecture. In this architecture, the frames are input to the network sequentially and in order, which facilitates the transfer of information of a frame to the previous and next frames. 
Another way is to use a batch of neighboring frames as the input to a feed-forward network for the reconstruction of a single frame (the middle frame in the batch). 
Each of these two approaches has its advantages and disadvantages. The networks in 
\cite{ResVSR, TGA, DUF, EDVR, TDAN}
 use the feed-forward structure with different initial feature extraction methods. In parallel, \cite{RBPN, BasicVSR, BasicVSRpp, FRVSR, BiRNN} use the RNN structure  to achieve acceptable results. \cite{tra1}, and \cite{tra2} have been able to improve the final accuracy by using transformers, at the expence of incorporating too many parameters and increasing the run-time.

In this paper, inspired by RCAN \cite{RCAN} we adopt the feed-forward structure with a residual in residual CNN; we further  improve its performance by the use of 
 grouped and depth-wise convolutional layers (which have much fewer computations). The later reduction in computational cost allows for increasing the number of network layers, which ultimately leads to a superior final accuracy.
  We further modify the batch normalization layers and employ 3D convolutional layers for processing adjacent frames.

\section{Related Work}

\begin{figure*}[ht]
	\centering
	\includegraphics[width=0.8\linewidth]{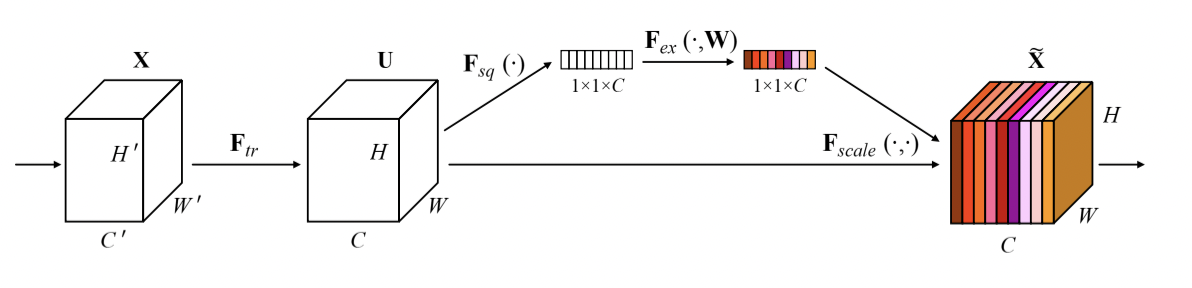}
	\caption{Channel attention mechanism. The image is adapted from \cite{SENet}.}
	\label{fig:SE}
\end{figure*}

\begin{figure*}[ht]
	\centering
	\includegraphics[width=0.5\linewidth]{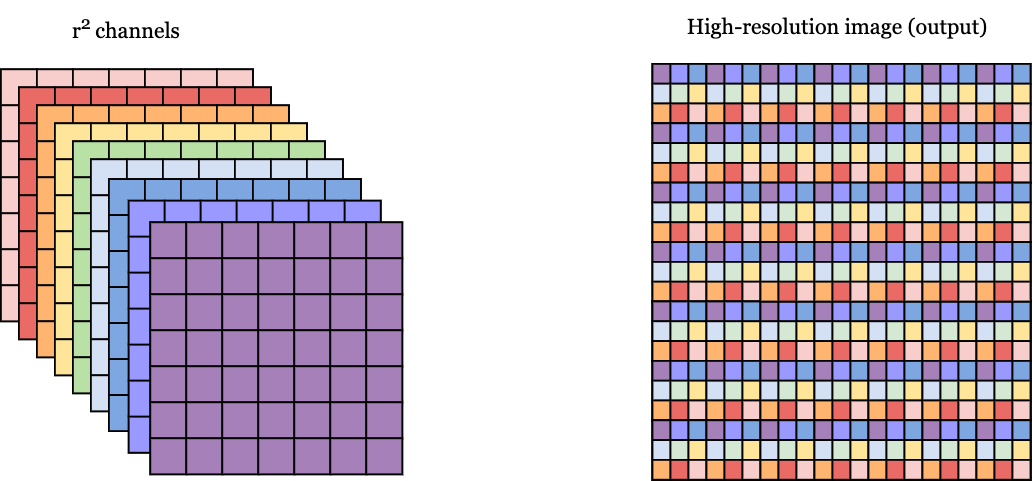}
	\caption{Pixel-shuffle method. By putting the corresponding pixels in $r^2$ channels together, we can enlarge the image in each dimension by a scale factor of $r$. The image is adapted from \cite{PixelShuffle}. }
	\label{fig:Pixel Shuffle}
\end{figure*}

\subsection*{Super-resolution}

A key difference between the CNNs for SR and other problems such as image classification is that the output is again an image. Because of this, the dimensions of the feature maps in the middle layers are large, and the computational cost of the network is very high. Due to upsampling at the beginning of the network and performing all the processing on the HR feature maps,  this issue is  more severe in SRCNN \cite{SRCNN}. Subsequent works including FSRCNN 
\cite{FSRCNN},
 improved on this by moving the upsampling block to the final layers of the network. More recent research works commonly 
 use the upsampling module in \cite{PixelShuffle} which has resolved the  issue of checkerboard artifacts 
 \cite{Check} caused by transposed convolutional layers in \cite{FSRCNN}. This method is depicted in Figure \ref{fig:Pixel Suffle}. Another problem with the structure of SRCNN \cite{SRCNN} is the small number of layers that limits the network capability.


While it is generally expected that by increasing the number of layers, better accuracies are achievable, simply stacking the convolutional layers could negatively affect the  training process, and is not generally a reliable solution. 
%
A similar problem is resolved in ResNet \cite{ResNet} by means of adding the input to the output of each block.
The advantages of this architecture have led most SR methods to use similar structures. Networks such as SRResNet \cite{SRGAN}, EDSR  \cite{EDSR},
and RCAN  \cite{RCAN} with residual block structures similar to ResNet \cite{ResNet}
 have yielded significant results in SISR. RCAN
\cite{RCAN} has made a step even further and has established the ability to create much deeper networks by adding long skip connections.

The main difference between SISR and VSR is the way we use  multiple input frames. The simplest forms the of recurrent neural networks use unidirectional propagation in which the information propagates from the previous frames to the next frames. FRVSR \cite{FRVSR} uses this  form.
In \cite{BiRNN}, bidirectional propagation is used where the network has two parts, direct and inverse. The obtained feature maps in both paths are put together to form the output.
BasicVSR \cite{BasicVSR} is among the examples that uses bidirectional propagation.
By modifying the way the information propagates, and in particular, the information transfer between the direct and reverse paths, IconVSR \cite{BasicVSR} and BasicVSR++ \cite{BasicVSRpp} have improved the results. 

The RNN structure also allows for taking advantage of the information in neighboring frames; nevertheless, 
 the complexity of the training process and problems such as vanishing or exploding gradients make it very difficult to enlarge the network structure. 
 Transformers can be used to resolve these issues and improve the performance of RNN structures. However, VSR models based on transformers usually involve many parameters to produce accurate outputs \cite{tra1}, \cite{tra2}.

A naive strategy to use the neighboring frames in a feed-forward architecture is to simply concatenate all of the frames as a single input to the network. 
The main drawback is that the temporal dependency between the frames is fully or partially lost. 
In other words, it is better to first preprocess the frames
and use the resulting feature maps as the input to the main part of the network.
 Networks in \cite{EDVR, TDAN, ResVSR} are of this type. Grouping the input frames \cite{TGA} and applying 3D convolutional layers\cite{DUF} are among two other ways to make use of the adjacent frames in feed-forward networks.

A common preprocessing block in most models is the motion compensation module between the frames. 
We remind  that our goal is to use the information in adjacent frames as much as possible. However, 
in different datasets and different scenes, the amount of motion between adjacent frames are unequal. The motion compensation module is responsible for dealing with this varying motion.
We should emphasize that, if the network structure is suitable and the training dataset is diverse enough, the motion compensation task shall be implicitly achieved in the network, without the need for any  preprocessing block. 
%
In case of including an explicit preprocessing motion compensation module, it should be very accurate and reliable \cite{DUF}. The two dominant types are those based on optical flow like \cite{RBPN, ToFlow}, and those based on deformable convolutional layers (these types of layers are introduced in \cite{Deformable}) like\cite{TDAN, EDVR, BasicVSR, BasicVSRpp}. 
Despite the use of explicit motion compensation methods in multiple works, these methods were ineffective in our structure and therefore, were not used.

\subsection*{Improving the performance of CNNs}

\begin{figure*}[ht]
     \centering
     \subfloat[The general structure of the network.]{
     \label{fig: mymodel1}
      \includegraphics[width=0.9\textwidth]{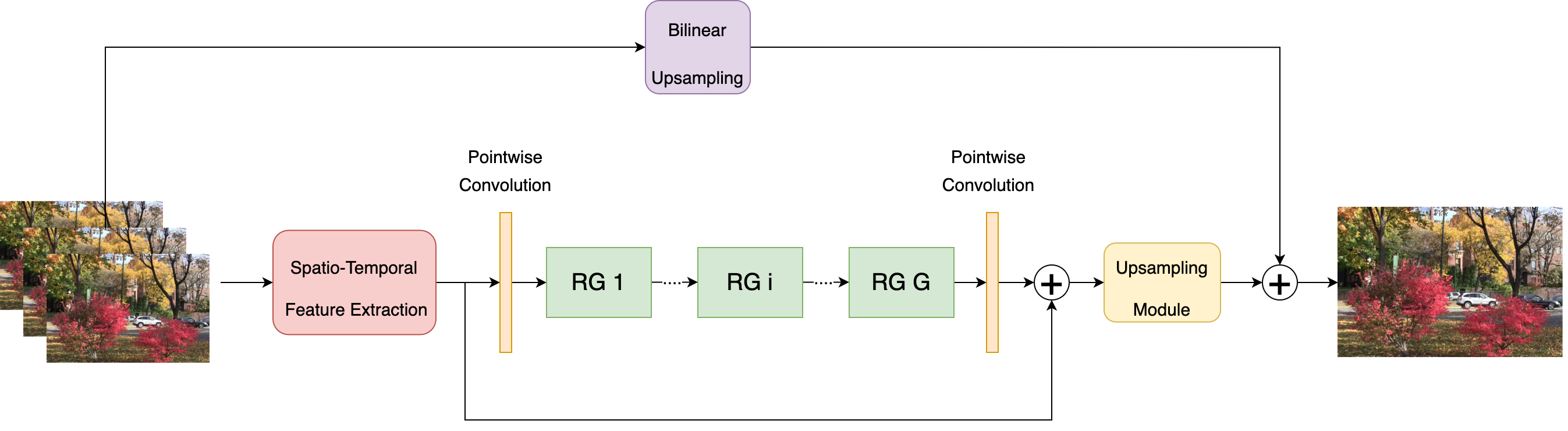}}

     \subfloat[The inner structure of residual groups.]{
     \label{fig: mymodel2}
      \includegraphics[width=0.5\textwidth]{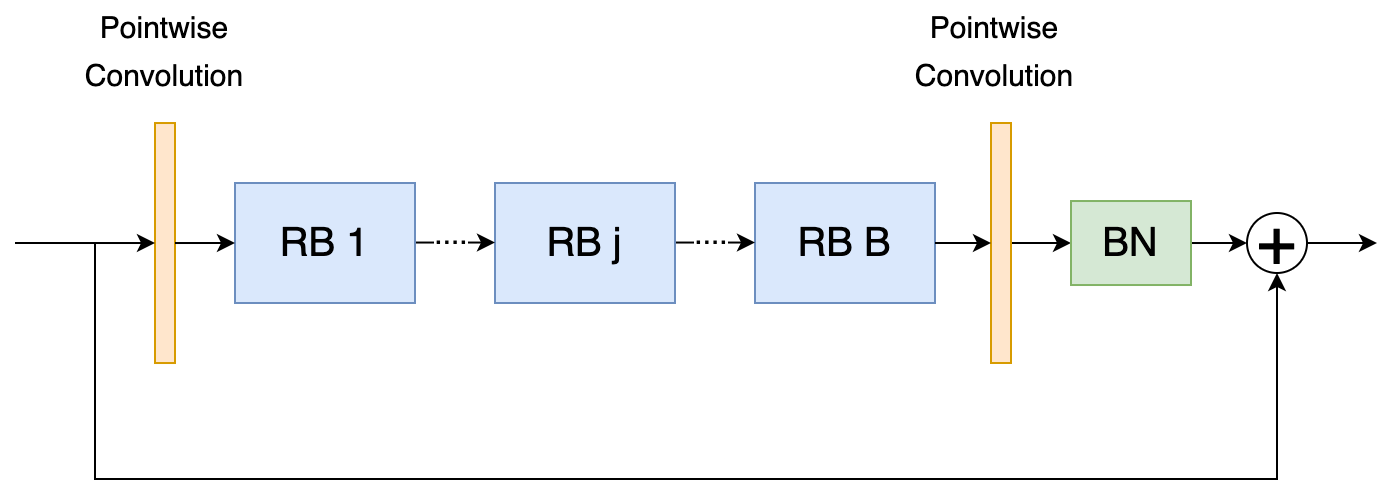}}

\caption{The structure of GRRN.}
\label{fig: mymodel}
\end{figure*}

There are several methods in the literature for improving the the performance of CNNs for the task of image classification. Fortunately, many of them are also applicable to the SR task.

The first method is to replace convolutional layers with grouped or depth-wise convolutional layers. 
In a normal convolutional layer, each output channel is a function of all input channels. Instead, in a grouped convolutional layer with $g$ groups, 
the number of parameters is almost reduced by a factor of $g$ without reducing the size of the feature map. We can even go a step further and put the value of $g$ equal to the number of channels to have a depth-wise convolutional layer. 
Models like MobileNet \cite{MobileNet},  ResNext \cite{ResNext}, and ShuffleNet \cite{ShuffleNet} have all benefited from this technique, and they all have relatively low computational cost as well as high output accuracy. 

Despite the desirable properties of grouped and depth-wise convolutional layers, we cannot restrict the architecture to be formed of solely this type of layers, as the network will be divided into unconnected groups from the beginning to the end.  To 
transfer information between such groups, $1 \times 1$ or point-wise convolutional layers are incorporated in ResNext \cite{ResNext} and MobileNet \cite{MobileNet}. The combination of depth-wise and point-wise convolution layers is called a separable convolution.
%
By replacing the convolutional layers with grouped/depth-wise and point-wise convolutional layers, the number of network parameters is significantly reduced, with a substantial portion of the processing power reserved for the point-wise convolutions.
In ShuffleNet \cite{ShuffleNet}, for further reducing the computational load, the point-wise convolutions are grouped. To transfer information between these groups, channel shuffling method is introduced.
 Using a  relatively similar method, we  create an efficient structure for the VSR challenge.

The channel attention mechanism used in \cite{SENet} is another successful technique to enhance the capability of CNNs.
%
In this mechanism, we assign more weight to the feature map channels that contain more important information.
For this purpose, one first applies average pooling to the feature map to construct a vector with the same length as the number of channels; this vector summarizes each feature map channel in a number.
Next, by passing this vector through two fully-connected layers, we obtain the weights. To avoid high computational loads in this attention mechanism, we can reduce the vector dimension by a factor $r$  after the first layer, which is then enlarged in the second layer.


Another common technique is to use batch normalization \cite{BN}, which expedites and stabilizes the training procedure in deep networks.
This technique consists of normalizing the activations in the middle layers.
In addition, by including two learnable parameters, we change the mean and variance of the output distribution.
In the training process, we use the empirical means and variances of minibatches for normalization, while in the test phase, as the data is no longer in form of minibatches, we use moving average statistics.

This issue is among the disadvantages of batch normalization which makes training setup somewhat different from the test setup. 
We shall provide a solution to this issue in the proposed structure.

\section{Grouped residual in residual network (GRRN)}

As shown in Figure \ref{fig: mymodel1}, the proposed model consists of three parts: spatio-temporal feature extraction, the main body, and the upsampling module. At the output, the middle input frame is bilinearly upsampled and added to the output of the upsampling module to produce the final image. This allows the main body of the network to focus on producing only the output details instead of the whole image. 

In the spatio-temporal feature extraction part, the features of the neighboring input frames are extracted using 3D convolutional layers. For the main body of the network, the core of RCAN
\cite{RCAN}
 is used, which means that the network has a residual in residual structure. This part consists of several residual groups and each group consists of several residual blocks. Important changes have been made to the structure of the residual blocks, including the use of separable convolutions and a change in the way batch normalization is used. These changes  enable us to achieve an efficient implementation. In the final part of the network, using the pixel-shuffle method
\cite{PixelShuffle} on the obtained feature maps, the details of the output image are reproduced and added to the raw image generated by the bilinearly-upsampled, to form the output.
 
Except for the fully-connected layers of the channel attention module, the PReLU nonlinear activity function is consistently used in all parts of the network. In this function, the slope of the line for the negative inputs is a learnable parameter  in the training process. This can prevent having dead zones such as in ReLU.

\subsection{Initial feature extraction}

\begin{figure*}[ht]
	\centering
	\includegraphics[width=0.9\linewidth]{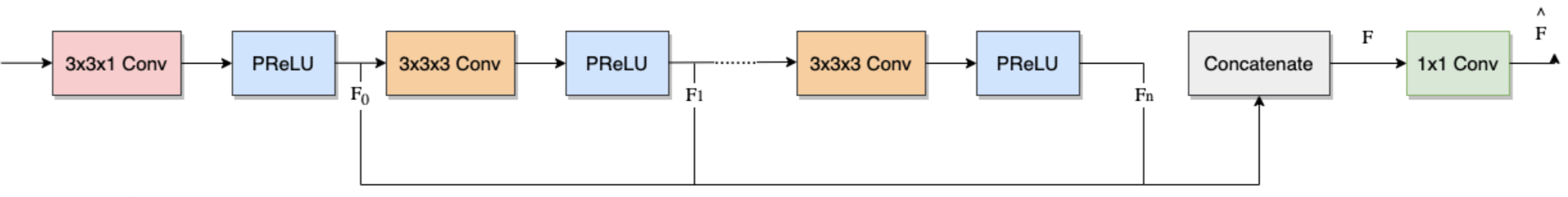}
	\caption{The structure of the initial feature extraction module.}
	\label{fig:mymodel3}
\end{figure*}

As shown in Figure \ref{fig:mymodel3}, we use 3D convolutional layers for feature extraction, as the network structure is feed-forward.
For the input, we include 
$2n+1$
low-resolution frames, namely the previous $n$ and the next $n$ frames apart from the main/current frame. 
In this work, we use $n=3$, which implies that the input consists of $7$ consecutive frames. 

Initially, we pass the 
frames through a three-dimensional $3 \times 3 \times 1$  convolutional layer, which is effectively a $3 \times 3$ 2D convolutional layer applied to all frames with shared parameters:
%
%
\begin{equation}
F_0^{H\times W\times (2n+1)\times s} = \delta \Big( C_{3 \times 3 \times 1} \big(I^{H\times W\times (2n+1)\times 3}\big) \Big)\ ,
\end{equation}
where $I$ is the set of input frames, $\delta$ is the PReLU nonlinear activity, and $C$ is a convolutional layer. Also, $H$ and $W$ are the dimensions of the input image, and $s$ is the number of output channels. The next layers use $3 \times 3 \times 3$ convolutional layers to extract appropriate spatio-temporal features from the input frames. These layers are applied with the \lq\lq same\rq\rq padding in the spatial domain and the \lq\lq valid\rq\rq  padding in the temporal domain. In other words, the spatial dimension of the output is equal to the input, but the temporal dimension decreases by two units in each step:
\begin{equation}
F_1^{H\times W\times (2n-1)\times s} = \delta \Big(C_{3 \times 3 \times 3} \big(F_0^{H\times W\times (2n+1)\times s} \big) \Big)
\end{equation}
By following this way for $n$ steps, we eventually reach the  temporal dimension of $1$, and extract the feature set $[F_0, F_1, ..., F_n]$ containing the spatio-temporal of the input frames.
By concatenating
$F_0, F_1, ..., F_n$
 and merging the dimensions of time and depth, a comprehensive feature map of the input frames is obtained. We also use a $1 \times 1$ convolutional layer to decrease the number of output channels to the appropriate number for the next step.
 \begin{align}
 F^{H\times W\times \big((n+1)^2\times s\big)}  = & \mbox{Reshape}\big(\mbox{Concatenate}(F_0, F_1,\cdots ,F_n)\big) , \nonumber \\
 {\widehat{F}}^{H\times W\times S}  = & C_{1\times 1} \big(F^{H\times W\times ((n+1)^2\times s)} \big).
\end{align}

\subsection{Main body of the network}

\begin{figure*}[ht]
	\centering
	\includegraphics[width=0.9\linewidth]{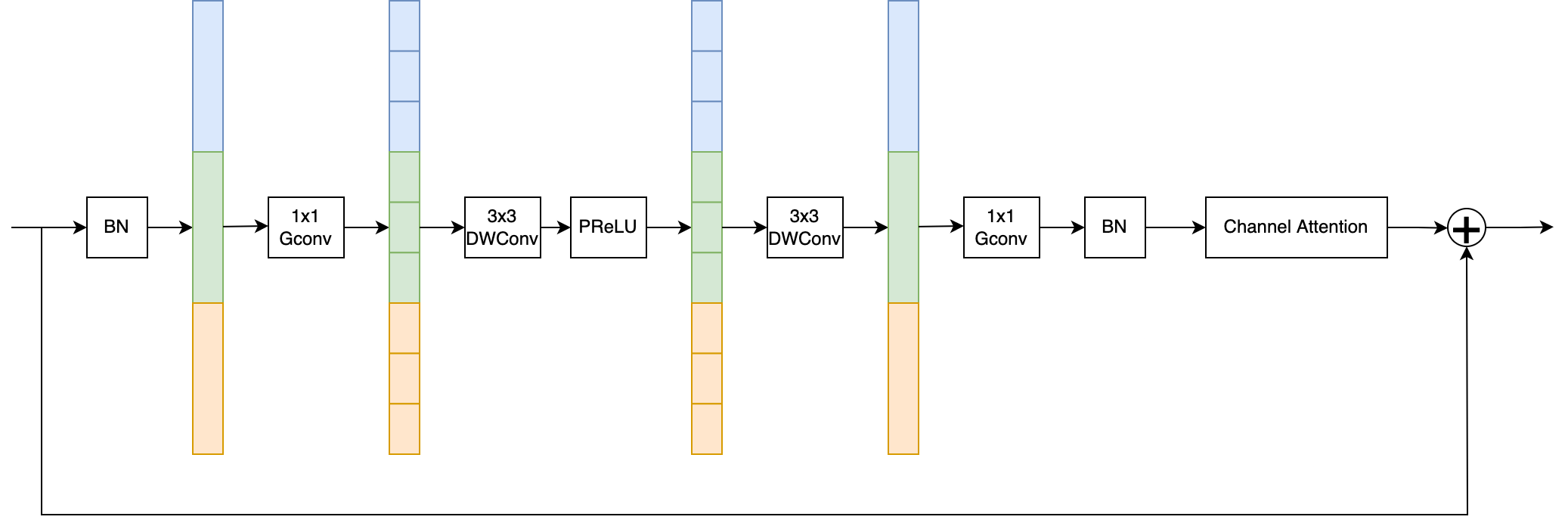}
	\caption{
	The inner structure of a residual block.}
	\label{fig:mymodel6}
\end{figure*}

The low-level features extracted in the previous part need to be converted into high-level features, from which we can reconstruct the details of the output SR  frame. 
As expected, 
this part of the network contains a large number of convolutional layers; as an example, 
ResNet-based structures \cite{ResNet}
 have shown successful results. 
Our proposed network also consists of residual blocks. Shortcut connections in this structure have two advantages: they make the training process smoother, and due to inclusion of the the block's input at its output, only a portion of residual details shall be reconstructed by each block.
In simple words, this approach is equivalent to reconstructing a high-resolution frame based on a low-resolution frame by successively incorporating the details from the feature maps. 
For the proposed network, a residual in residual structure similar to RCAN
\cite{RCAN} is designed. Long shortcut connections in this structure help in training the network 
despite having a huge number of layers. They also facilitate the flow of information in the network. Figure 
\ref{fig: mymodel}
 shows the general network structure and the inner structure of the residual groups. Next, we shall explain the involved components. 

\subsubsection*{Efficient convolutional layers}
In the proposed network, unlike RCAN \cite{RCAN}, we use grouped and depth-wise convolutional layers.
Such layers were previously implemented in ResNext \cite{ResNext}, MobileNet \cite{MobileNet}, and ShuffleNet \cite{ShuffleNet} for the task of image classification with favorable outcome. The point-wise $(1 \times 1)$ convolutional layers also play an important role in these types of structures. 
Unfortunately, grouped and depth-wise convolutional layers are mainly ignored for the SR problem.
  
The proposed model consists of a number of blocks. Besides the activation function, the batch normalization layers, and the channel attention segment, each block consists of a grouped point-wise convolutional layer at the beginning, a grouped point-wise convolutional layer at the end, and two depth-wise convolutional layers  in the middle. In the blocks, except for the calculation of the weights in the channel attention module, the parts of the feature maps (corresponding to the grouping of the point-wise convolutional layers) are processed separately (they are not mixed). We further form \emph{groups} by a number of consecutive blocks, a point-wise convolutional layer before the blocks, a point-wise convolutional layer after the blocks, and a batch normalization layer at the end. These point-wise convolutional layers mix the information in the isolated parts of the feature maps. 
%
%
This structure is found to be more efficient than those used in ResNext \cite{ResNext} and MobileNet \cite{MobileNet} which employ simple point-wise convolutional layers in each block; the grouped point-wise convolutional layers consist of only a fraction  of the parameters present in simple point-wise convolutional layers.


\subsubsection*{Channel attention mechanism}

As explained earlier, the channel attention mechanism improves the capabilities of CNN networks.
As shown in Figure 
\ref{fig:mymodel6}, the channel attention module is used at the end of each block in our architecture. This module can be thought of as a gateway for passing  information proportional to the significance of that information. 
 The quantitative effect of this method on the final results is investigated in Section \ref{sec:Ablation}.

\subsubsection*{Improved batch normalization}

\begin{figure*}[t]
	\centering
	\includegraphics[width=1\linewidth]{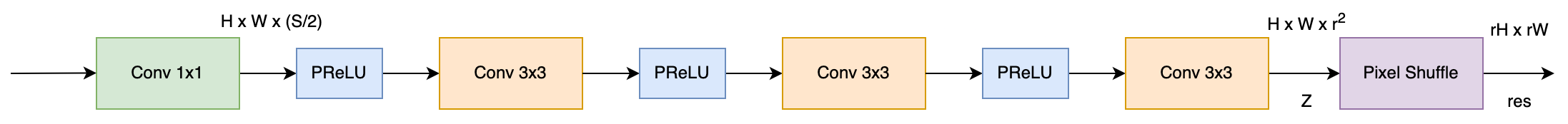}
	\caption{The inner structure of the upsampling module.}
	\label{fig:mymodelup}
\end{figure*}

To take advantage of the positive effects of batch normalization, we use this layer  at the beginning of each residual block, at the end of each residual block, and at the end of each residual group (Figures \ref{fig: mymodel2} and \ref{fig:mymodel6}). The numerical results  
show that the effect of batch normalization at the beginning of the training process is very positive and the network parameters quickly converge to fair values. 
In addition, a relatively large initial learning rate does not cause instability in the training process.
Nevertheless, the major drawback of batch normalization 
 is the mismatch between the training and testing scenarios.
 In particular, the empirical mean and variance of a minibatch are used during training, while the moving average of statistics is used during the tests, 
 which includes statistical information from a much larger part of the training dataset. The effect of this statistical difference 
 is even exacerbated by deepening the network. This is especially important when the minibatch size is small. In our experiments with small minibatch sizes, we even observed instability in the training process (as the calculated statistics on different processing cores are not usually synchronized). 
  In \cite{Renorm}, batch renormalization is proposed as a remedy. With the correction parameters introduced in this method, the overall estimated statistical parameters of the training process are made as close as possible to the moving average statistics. 
  At first, the effect of this correction is very small, but gradually increases during the training process, 
  and after a while, causes these layers to work almost identically in the training and testing phases. 
  In other words, the batch normalization layer positively influences the  training process at the early stages, while its drawbacks are avoided later.

Inspired by this method, we train batch normalization layers in few initial epochs of our network to increase the convergence rate at the beginning. Next, we keep batch normalization parameters fixed to avoid the negative impacts on the training process and to provide uniformity between the training and testing phases.

\subsection{Upsampling module}
In the main body of the network, the dimensions of all feature maps are equal to the size of the low-resolution input frames. It helps us to keep the computational cost of the network at a minimal level; however, the high-resolution frame shall be reconstructed at some stage, which necessitates an upsampling operation. We postpone this operation to the 
end of the network. 
%
%
As shown in Figure \ref{fig:mymodelup}, the feature map obtained from the main body of the network with depth $S$, is first compressed using a point-wise convolutional layer, and then, passed through $3$ layers of convolution, resulting in a feature map with  $r^2$ channels, where  $r$ is the magnification ratio of SR. To produce the final image, we employ the pixel-shuffle method \cite{PixelShuffle} for putting the pixels of different channels together:
\begin{align}
& {\rm res}_ {rx+l, ry+k} = Z_ {x,y,rl+k} , \nonumber\\
 x\in [0:H-1],\ & y\in [0:W-1],\ l,k\in [0:r-1],
\end{align}
where 
$Z^{H\times W\times {r^2}}$
represents the feature map obtained in the last layer, and 
${\rm res}^{{rH}\times {rW}}$
is the residual value of the output image.

\begin{table}[t]
\centering
\caption{The hyperparameters of the three implemented networks. The numbers in the parentheses are for batch normalization parameters which are kept fixed after initial epochs.}

\begin{tabular}{|l|c|c|c|c|c|c|c|}
\hline
& s
& S
& B
& G
& g
& r
& \begin{tabular}[c]{@{}c@{}}Learnable \\ parameters (M)\end{tabular}
\\ \hline 
GRRN-S
& 12 
& 192 
& 20 
& 4
& 3
& 32
& 3.06 + (0.06)
\\ \hline
GRRN
& 24
& 256
& 30 
& 6
& 4
& 32 
& 8.94 + (0.19) 
\\ \hline
GRRN-L
& 24
& 256
& 30
& 11
& 4
& 32
& 16.05 + (0.34)
\\ \hline
\end{tabular}
\label{tab: Hyper}
\end{table}

\begin{table*}[t]
\caption{Quantitative results of different methods on Vimeo90k-T and vid4. The results for ToFlow
\cite{ToFlow}
 are quoted from 
\cite{BasicVSRpp}.
 }
 
\resizebox{\textwidth}{!}{\begin{tabular}{|c|c|c|c|c|c|c|c|c|c|c|c|}
\hline

&Bicubic
&ToFlow \cite{ToFlow}

&RBPN \cite{RBPN}

&EDVR \cite{EDVR}

&BasicVSR \cite{BasicVSR}

&IconVSR \cite{BasicVSR}

&BasicVSR++ \cite{BasicVSRpp}

&GRRN-S
&GRRN
&GRRN-L
\\ \hline 
Vimeo90k-T-PSNR
&31.26
&33.08 
&37.07 
&37.61 
&37.18 
&37.47 
&\textcolor{red}{37.79} 
&37.13 
&37.48 
&\textcolor{blue}{37.67} 
\\ \hline 
Vimeo90k-T-SSIM
&0.8683 
&0.9054 
&0.9435 
&0.9489 
&0.9450 
&0.9476 
&0.9500 
&0.9478 
&\textcolor{blue}{0.9511} 
&\textcolor{red}{0.9513} 
\\ \hline
vid4-PSNR
&23.78 
&25.89 
&27.12 
&27.35 
&27.24 
&27.39 
&\textcolor{red}{27.79} 
&27.21 
&27.36 
&\textcolor{blue}{$27.47$} 
\\ \hline 
vid4-SSIM
&$0.6347$ 
&$0.7651$ 
&$0.8180$ 
&$0.8264$ 
&$0.8251$ 
&$0.8279$ 
&\textcolor{red}{$0.8400$} 
&$0.8222$ 
&$0.8271$ 
&\textcolor{blue}{$0.8308$} 
\\ \hline
\begin{tabular}[c]{@{}c@{}}Learnable \\ parameters (M)\end{tabular}
&$-$
&$-$
&$12.2$
&$20.6$
&$6.3$
&$8.7$
&$7.3$
&$3.06+(0.06)$
&$8.94+(0.19)$
&$16.05+(0.34)$
\\ \hline
\end{tabular}}
\label{tab: Scores1}
\end{table*}

\begin{figure*}[ht]

\begin{tabular}{ccccc}
      \small
      
	  \multirow{1}{*} {\includegraphics[width=0.33\linewidth]{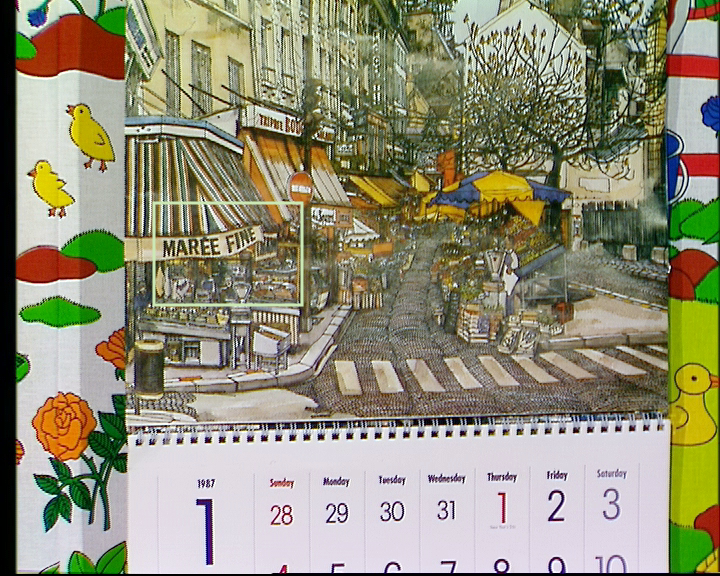}} &  Bicubic & BasicVSR & IconVSR  &  BasicVSR++ \\&
      \addheight{\includegraphics[width=0.15\linewidth]{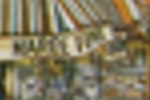}} &
      \addheight{\includegraphics[width=0.15\linewidth]{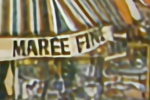}} &
      \addheight{\includegraphics[width=0.15\linewidth]{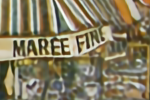}} &
      \addheight{\includegraphics[width=0.15\linewidth]{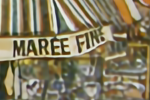}} 
      \\
      	 & GRRN-S  &  GRRN & GRRN-L & HR \\&
      \addheight{\includegraphics[width=0.15\linewidth]{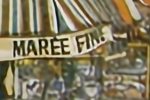}} &
      \addheight{\includegraphics[width=0.15\linewidth]{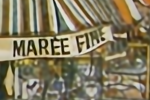}} &
      \addheight{\includegraphics[width=0.15\linewidth]{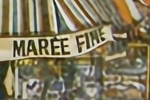}} &
      \addheight{\includegraphics[width=0.15\linewidth]{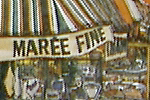}} \\
\end{tabular}

\begin{tabular}{ccccc}
      \small
      	  \multirow{1}{*}{\includegraphics[width=0.32\linewidth]{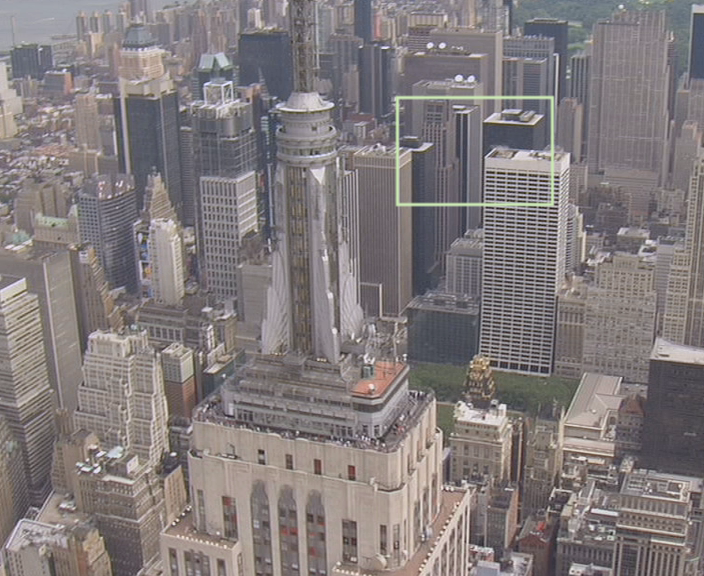}} &  Bicubic & BasicVSR & IconVSR  &  BasicVSR++ \\&

      \addheight{\includegraphics[width=0.15\linewidth]{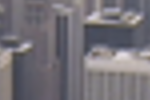}} &
      \addheight{\includegraphics[width=0.15\linewidth]{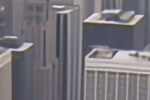}} &
      \addheight{\includegraphics[width=0.15\linewidth]{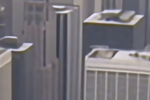}} &
      \addheight{\includegraphics[width=0.15\linewidth]{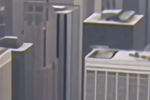}} 
      
      \\
      	   & GRRN-S  &  GRRN & GRRN-L & HR \\&
      \addheight{\includegraphics[width=0.15\linewidth]{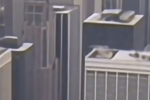}} &
      \addheight{\includegraphics[width=0.15\linewidth]{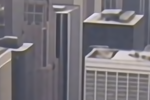}} &
      \addheight{\includegraphics[width=0.15\linewidth]{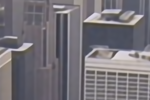}} &
      \addheight{\includegraphics[width=0.15\linewidth]{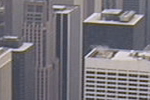}} \\
\end{tabular}

\caption{Outputs of different models on sample input frames.
}
\label{fig: outputs}
\end{figure*}

\section{Experiments}


The main dataset used in our experiments is Vimeo90K 
\cite{ToFlow} which is the most common dataset for benchmarking VSR techniques. This dataset includes $91401$ video sequences each with  $7$ frames. Among these video sequences, $64612$ are for the training set, $7824$ for the test set, and the remaining $19265$ can be used for the validation set.
The resolution of the low-resolution (input) frames is $64\times 112$ and the desired high-resolution frames are with the size $256 \times 448$. Actually, the low-resolution frames are generated by  downsampling the high-resolution frames using the bicubic method. 

The vid4 dataset has also been used as another test dataset, which contains only $4$ video sequences ranging from $30$ to $50$ frames each. 
The spatial resolution of this dataset is larger than Vimeo90K. 

Based on the proposed structure, three networks with different numbers of learnable parameters have been implemented.  The hyperparameters in this structure are: i) the depth $s$ of the feature maps in the spatio-temporal feature extraction part, ii) the depth $S$ of feature maps in the main body of the network, iii) the number of residual blocks $B$ in each residual group, iv) the total number $G$ of residual groups, v) the number of pointwise convolution groups $g$ in each residual block, and vi) the ratio $r$ of reduction in channel attention mechanism. Table 
\ref{tab: Hyper} summarizes these hyperparameters for the three implemented models.
 
The pixels in the input and output frames 
are on an $8$-bit scale ($0$ to $255$). Because the main body of the network produces only the residual value of the output (which has a much smaller dynamic range than the input frames) frames are scaled by a factor of 
$\frac{1}{25.5}$ at the beginning of the network. Also, to make the training process more stable, 
small initial values for  the coefficient $\gamma$ (a learnable parameter in the batch normalization layer) have been selected at the end of the groups and blocks. 

The network is trained using a cloud TPU. This hardware is optimal for tensor processing and significantly improves the training speed. Charbonnier loss \cite{Charbonnier} is set as the loss function, Minibatch size is tuned to $16$ and Adam optimizer with an initial learning rate of $0.0004$ is employed. In certain epochs, the learning rate is halved until finally
$\frac{1}{32}$ of the initial learning rate is reached. Also, after the first $5$ epochs, the batch normalization parameters  are kept constant.

\subsection{VSR quality comparison}

We compare our proposed method(s) with some of the existing techniques on Vimeo90k-T and vid4 datasets based on the PSNR and SSIM metrics.
Table \ref{tab: Scores1} summarizes the results. 
Interestingly, 
the implemented models provide competitive results; in particular, the GRRN-S method with only a fraction of RBPN
\cite{RBPN}
parameters provides even better results. The GRRN and GRRN-L methods yield the best performance 
on the Vimeo90k dataset in terms of the SSIM criteria. 
the BasicVSR++ method \cite{BasicVSRpp}, however, provides the best performance based on the PSNR metric (both datasets) and even based on the SSIM metric on the vid4 dataset.
We should highlight that with the vid4 dataset, the architecture of BasicVSR++ allows for including all the frames in a video ($30$ to $50$ frames) to produce a single frame, while the feed-forward networks only use $7$ consecutive frames. 

Figure 
\ref{fig: outputs}
 shows the outputs of different methods for two different sets of inputs. In these samples, our method performs better than BasicVSR++\cite{BasicVSRpp} (e.g., focus on the windows of the shorter building in the second sample).

\subsection{Test augmentation}

\begin{table}
\centering
\caption{The effect of test augmentation on the performance of the three networks on vid4 dataset. }
\begin{tabular}{|r|c|c|}
\hline
&
w/o test augmentation
&
w/ test augmentation
\\ \hline 
GRRN-S
&$27.21/0.8222$
&$27.32/0.8250$
\\ \hline
GRRN
&$27.36/0.8271$
&$27.48/0.8299$
\\ \hline
GRRN-L
&$27.47/0.8308$
&$27.58/0.8334$
\\ \hline
\end{tabular}
\label{tab: Aug}
\end{table}

Although our networks are trained on a large number of images/frames, there is no gaurantee that the network becomes flip- or rotation-invariant. 
For instance, if we horizontally flip all the input frames, the output frames  are not necessarily flipped. 
We believe that making the algorithm symmetric (flip- and rotation-invariant), improves the VSR quality. For this purpose, similar to \cite{EDVR}, we apply a transformation to the input frames and apply  its inverse to the output.
Due to the asymmetry of the network, this output differs from the one achieved without applying any transformation. We consider $7$ simple transformations including horizontal and vertical flips, rotation, and their combinations; finally, we average all the $8$ results (one without any transformation and $7$ with transformations).
 Table 
\ref{tab: Aug}
 shows the effect of applying this method to the three implemented models. 

\subsection{Ablation study}\label{sec:Ablation}

In this section, we test the effect of some network components on the performance. The tests are done on the GRRN-S model.

\subsubsection{The effect of the residual in residual structure}

\begin{table}
\centering
\caption{ The effect of removing the residual in residual structure on the results of GRRN-S model on Vimeo90k-T (to keep the number of parameters the same, the total number of residual blocks is increased from $80$ to $90$).}
\begin{tabular}{|c|c|}

\hline
simple residual structure
&
residual in residual structure
\\ \hline 
$36.89/0.9455$
&$37.13/0.9478$
\\ \hline
\end{tabular}
\label{tab: RIR}
\end{table}

As explained before, the groups of blocks have two important properties:
\begin{itemize}
\item
Point-wise convolutional layers at the beginning and at the end of each  group allow the transfer of information between different parts of the feature maps in the blocks (corresponding to the grouped depth-wise convolutional layers).

\item
Shortcuts from the beginning to the end of each  group improve the network training process and allow for increasing the total number of layers.
\end{itemize}

Table
\ref{tab: RIR} confirms that removing the residual in residual structure has a degrading effect on the results.

\subsubsection{The effect of the channel attention mechanism}

\begin{table}
\centering
\caption{ The effect of removing the channel attention mechanism on the results of GRRN-S model on Vimeo90k-T.}
\begin{tabular}{|c|c|}

\hline
\begin{tabular}[c]{@{}c@{}}w/o channel \\ attention mechanism
\end{tabular}
 
&
\begin{tabular}[c]{@{}c@{}}w/ channel \\ attention mechanism
\end{tabular}

\\ \hline 
$37.08/0.9463$
&$37.13/0.9478$
\\ \hline
\end{tabular}
\label{tab: SE}
\end{table}

\noindent
Table 
\ref{tab: SE}
 shows the  results of GRRN-S with and without using the channel attention mechanism. The results identify the channel attention mechanism as a useful module.

\subsubsection{The effect of batch normalization}

\begin{table}[t]
\centering
\caption{ The effect of freezing batch normalization after the first five epochs. The test dataset is Vimeo90k-T, and the minibatch size is 64.}
\begin{tabular}{|c|c|}

\hline
BN training continues
&
BN training stops
\\ \hline 
$36.84/0.9456$
&$36.87/0.9464$
\\ \hline
\end{tabular}
\label{tab: BN}
\end{table}

Due to the large size of the network and the training dataset, it is very time-consuming to train the network on a single processing core. This necessitates  parallelizing the computations. However, in the absence of consistent calculation of the statistics in different processing cores, the problem of differing statistics can be exacerbated. 
We mentioned earlier that if the size of the minibatch is small, it is  likely to observe instability in the training process. For this purpose, after few initial epochs, we keep the parameters of the batch normalization layers constant. Here, we show that whether the minibatch size is small or not, freezing the batch normalization layers after the initial epochs does not have a negative effect (actually  a very positive effect for small minibatch sizes).
%
For the minibatch size of $64$,
Table 
\ref{tab: BN} 
shows 
that the proposed method still improves the final performance. In other words, we observe that the strength of the batch normalization is mainly limited to the early stages of the training.

\section{Conclusion}

We proposed a grouped residual in residual network (GRRN) for the purpose of video super-resolution. A structure made up of residual blocks and residual groups allows us to build very deep networks. We made this structure more efficient by using point-wise, grouped, and depth-wise convolutional layers. Also, the use of a channel attention mechanism in residual groups slightly increased the network capacity. We further proposed a special way to take advantage of batch normalization layers. The proposed network is compared with the existing methods for VSR and demonstrated acceptable performances.

\bibliographystyle{IEEEtran}
\bibliography{Citations.bib}

\end{document}